\documentclass[aps,pra,twocolumn,showpacs,epsfig]{revtex4}
\usepackage{dcolumn}
\usepackage{bm}
\usepackage{graphicx}
\usepackage{amsmath}
\usepackage{latexsym}
\usepackage{amsfonts}
\usepackage{amssymb}
\usepackage{array}
\usepackage{times}
\usepackage{epsfig}
\usepackage{epstopdf}
\usepackage{setspace}

\newcommand{\ket}[1]{\left\vert#1\right\rangle}

\newcommand{\bra}[1]{\left\langle#1\right\vert}

\newcommand{\nbar}{\overline{n}}

\def\be{\begin{equation}}
\def\ee{\end{equation}}
\def\bea{\begin{eqnarray}}
\def\eea{\end{eqnarray}}
\def\beas{\begin{eqnarray*}}
\def\eeas{\end{eqnarray*}}
\def\l({\left(}
\def\r){\right)}
\def\l[{\left[}
\def\r]{\right]}

\def\f{\frac}

\def\ba{\begin{array}}
\def\ea{\end{array}}

\begin{document}
\title{Genuine multipartite nonlocality of entangled thermal states }
\author{
G. McKeown$^1$, F. L. Semi\~ao$^2$, H. Jeong$^{3}$, and M. Paternostro$^1$}
\affiliation{$^1$School of Mathematics and Physics, The Queen's University, Belfast,
BT7 1NN, UK\\
$^2$Departamento de F\'isica, Universidade Estadual de Ponta Grossa- Campus Uvaranas, 84030-900, Ponta Grossa, Paran\'{a}, Brasil\\
$^3$Department of Physics and Astronomy, Seoul National University,
 Seoul, 151-742, Korea}
\date{\today}
\begin{abstract}

We assess quantum non-locality of multiparty entangled thermal states by studying, quantitatively, both tripartite and quadripartite states belonging to the Greenberger-Horne-Zeilinger (GHZ), W and linear cluster-state classes and showing violation of relevant Bell-like inequalities. We discuss the conditions for maximizing the degree of violation against the local thermal character of the states and the  inefficiency of the detection apparatuses. We demonstrate that such classes of multipartite entangled states can be made to last quite significantly, notwithstanding adverse operating conditions. This opens up the possibility for coherent exploitation of multipartite quantum channels made out of entangled thermal states. Our study is accompanied by a detailed description of possible generation schemes for the states analyzed.
\end{abstract}
\pacs{03.67.Mn, 42.50.Dv, 03.65.Ud, 42.50.-p}

\maketitle

\section{Introduction}

Where lays the boundary between classical and quantum worlds? The daunting nature of this question is not preventing the pursuit of very interesting studies digging deeply into the origin of non-classicality of a physical system. On the contrary, the difficulties inherent in such a fundamental investigation are  sharp stimuli to the research of tests and physical systems able to challenge the common belief that quantumness occurs only under quite {\it special}, and yet hard to achieve, conditions. Very recently, quite significant endeavors have been produced in the study of quantumness at the ``large scale" by proposing ways to infer non-classicality in multi-photon states and massive mechanical oscillators~\cite{Martini,Martini2,Martini3,Martini4,branciard,kipp,kubala}. 

In this article we contribute to such a quest by addressing the noteworthy case of multipartite quantum correlations shared by systems that, when individually taken, are fully classical and revealed by instruments far from offering any single-quantum resolution. We address the case of multipartite entanglement shared by bosonic systems that are locally prepared in chaotic thermal states, which are commonly intended as well-defined classical entities, and dub them entangled thermal states (ETS)~\cite{jr06}.  We demonstrate that such quantum correlations can be easily made strong enough to be revealed, through the violation of suitable Bell-like inequalities, against any initial local temperature and regardless of the coarse-graininess of the detectors used in order to implement the non-locality test. We first study ETS versions of tripartite GHZ~\cite{ghz} and W states~\cite{Wstates}, which are prominent and non-equivalent classes of three-particle entangled states. We discuss two schemes for generating GHZ-like ETS and draw a comparison between the slightly different states therefore obtained. We show that an inequality for genuine multipartite non-locality can be violated up to the maximum value allowed for a given representative of the class of GHZ-like ETS, when effective local rotations and highly inefficient and noisy homodyne measurements are employed. Similar action can be taken for the ETS version of W states~\cite{Wstates}, where the violation is not to the maximum degree. We then extend our analysis to larger entangled states, proving independence of a few of our results from the number of bosonic systems involved in the multipartite states we scrutinize. Such larger ETS states include the quadripartite version of GHZ-like ETS and a four-mode linear cluster-like ETS. Our results go along and are consistent with a recently-started line of investigation aiming at showing that quantumness can be enforced and revealed in situations that are at the verge of the classical world~\cite{jr06,jacobmaurotim,branciard}. 

The remainder of this paper is organized as follows. In Sec.~\ref{tripartite} we assess genuine tripartite non-locality in GHZ and W-like ETS. The main tool of our investigation is an inequality derived by Svetlichny~\cite{svetl}. We sketch two schemes for the generation of GHZ-like ETS states and provide an operative protocol for testing tripartite non-locality by means of effective rotations and arbitrarily efficient homodyne measurements. In Sec.~\ref{quadripartite} we tackle the non-local properties of four-party states such as quadripartite GHZ-like ETS and the interesting class of linear cluster-like ETS. Non-locality can be revealed regardless of the local temperature and can be made robust to the inefficiency and fuzziness of the detection devices. Finally, Sec.~\ref{conclu} summarizes our findings. 


\section{Tripartite case: of GHZ and W-like ETS}
\label{tripartite} 

\begin{figure*}[ht]
\center{\psfig{figure=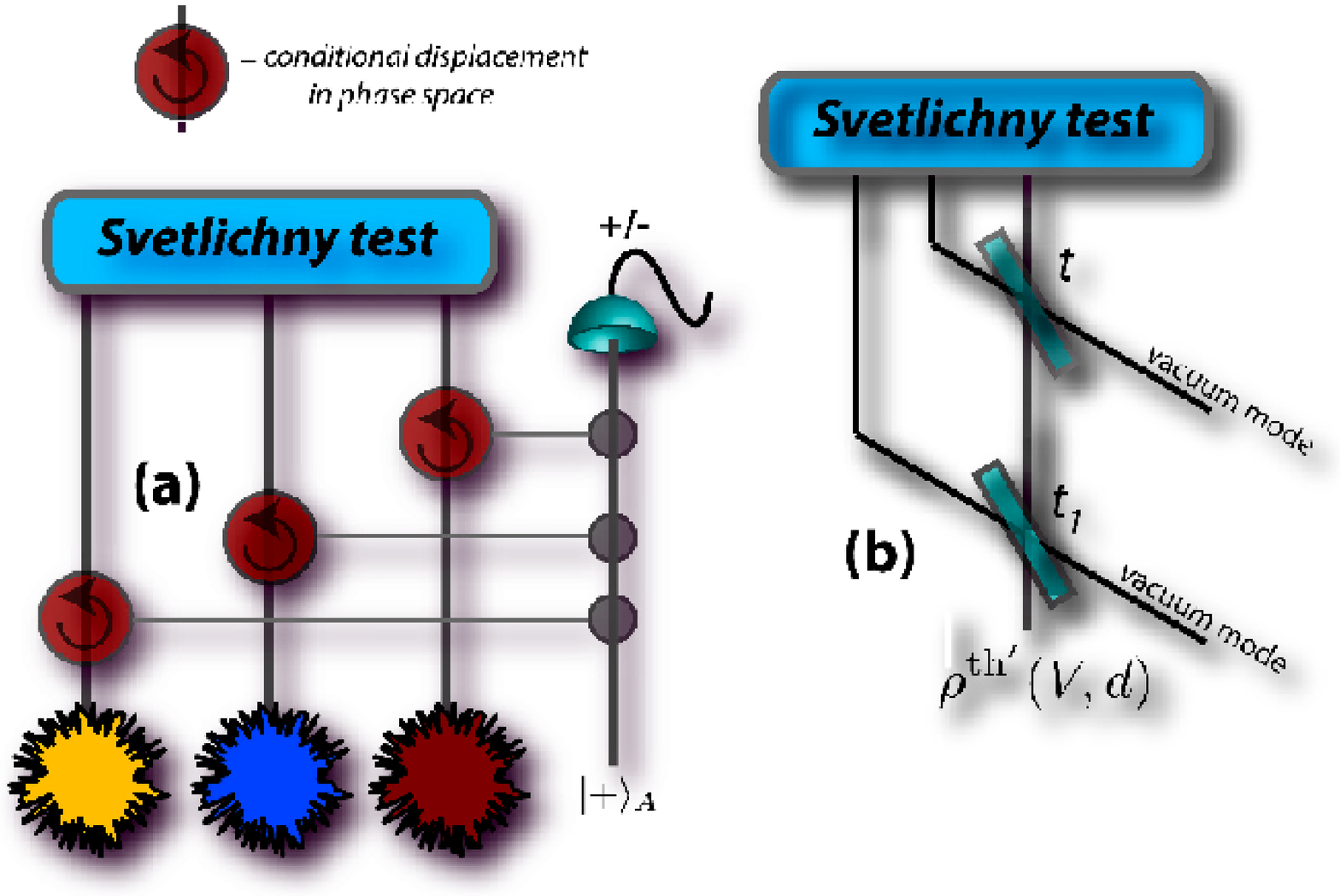,width=7cm,height=4.5cm}\hskip1.5cm\psfig{figure=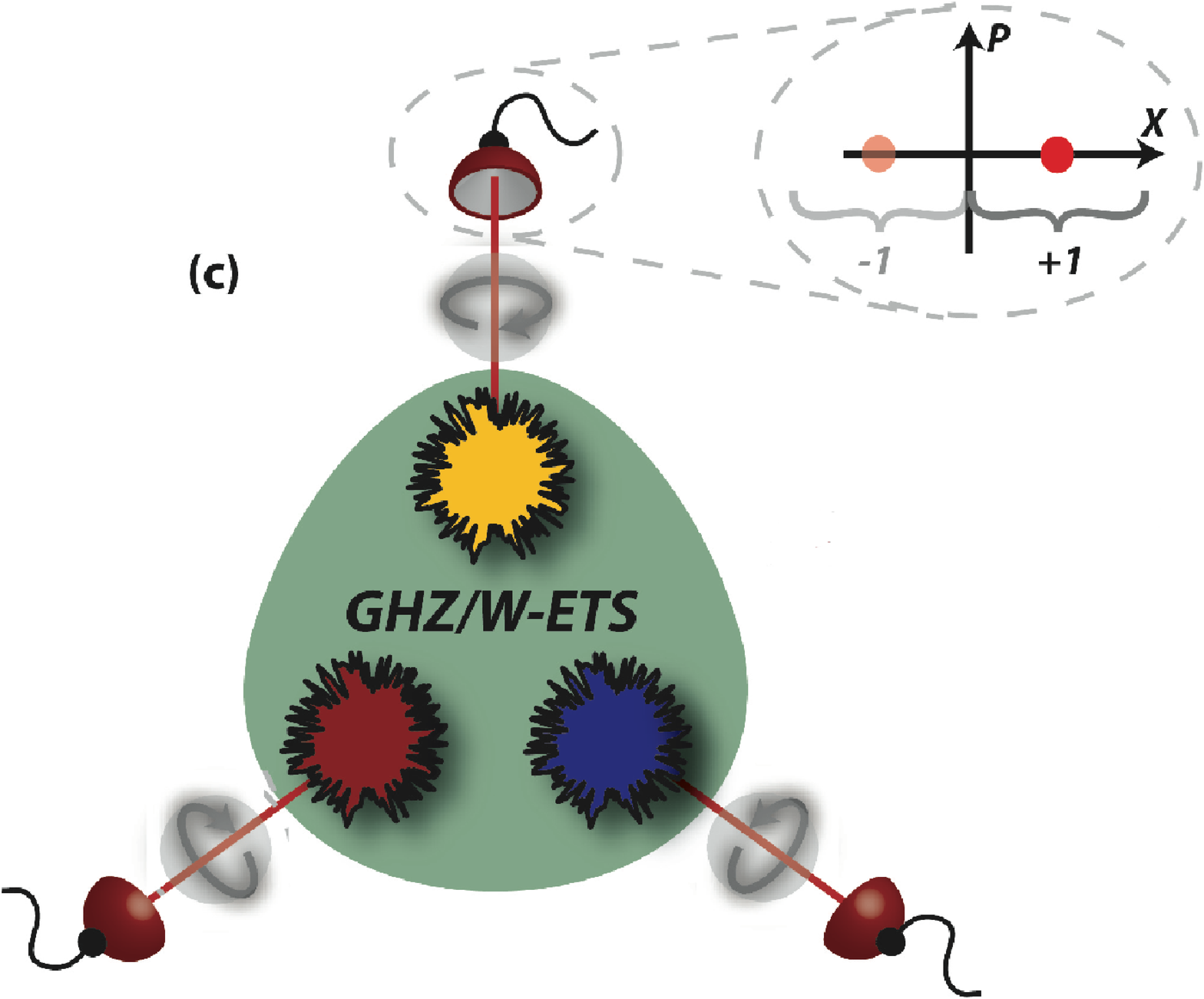,width=7cm,height=5cm}}
\caption{{\bf (a)} Conditional scheme for generating GHZ-like ETS. Three displaced thermal states (labelled $j=1,2,3$) interact with a two-level ancilla prepared in state $\ket{+}_A=(\ket{0}_A+\ket{1}_A)/{\sqrt{2}}$. The interaction is ruled by the coupling Hamiltonian $\hat{H}_{Aj}=\hbar\Omega|1\rangle_A\langle{1}|\hat{a}^\dag_j\hat{a}_j$ for a time equal to $\pi/\Omega$. The ancilla is finally projected onto $\ket{\pm}_A$ with $_{A}\langle+|-\rangle_A=0$. {\bf (b)} Beam splitter-based scheme for the generation of tripartite GHZ-like ETS. A displaced thermal state as introduced in the body of the paper undergoes a transformation that puts it into the form $\rho^{\text{th}'}(V,d)={\cal N}_+\int{d}^2\alpha P^{\text{th}}_\alpha(V,d)\ket{s^+_\alpha}\bra{s^+_\alpha}$ with $\ket{s^+_\alpha}$ defined as in Eq.~(\ref{deff}) and ${\cal N}_+$ a normalization constant. This is then superimposed to two vacuum modes at beam splitters of transmittivity $t_1=\sqrt{2/3}$ and $t_2=1/\sqrt{2}$~\cite{jeongNguyen}. {\bf (c)} Scheme for Svetlichny test performed over a GHZ/W-like ETS. Each mode of a state prepared off-line is first appropriately rotated and then projected onto in-phase quadrature eigenstates by means of arbitrarily efficient homodyne detectors. The outcomes of the measurements are appropriately dichotomized (see the other inset).}
\label{generazione}
\end{figure*}

\subsection{Svetlichny versus Mermin inequality for spin-$1/2$ particles}

The quantification of the degree of genuine multipartite entanglement in the aforementioned class of tripartite and quadripartite ETS is quite a daunting task. For pure states of three entangled qubits, it is possible to quantify the degree of multipartite entanglement~\cite{KKW}. However, this is not the case for the class of ETS here at hand. In fact, as it will be seen in Sec.~\ref{gen}, we will be dealing with highly mixed states of systems spanning  infinite dimensional Hilbert spaces. Moreover, by construction, such states do not belong to the class of so-called Gaussian states, {\it i.e.} states whose Wigner function is a Gaussian in phase-space~\cite{BR}. While correlations in Gaussian states are well and easily characterized, we face the
lack of necessary and sufficient criteria for the quantification of entanglement in non-Gaussian states. In fact, the available entanglement measures in continuous variable are based (to the best of our knowledge) on the use of the negativity of partial transposition criterion formulated in terms of covariance matrices, which carry exact information on the state of a system only in the Gaussian scenario. Therefore, even the case of quantifying bipartite entanglement represents quite a challenging problem. A partial assessment of the entanglement content of a given two-mode state belonging to such class passes through so-called Òentangling powerÓ methods: one could establish a lower bound to the entanglement within a two-mode non-Gaussian mixed state by determining the entanglement that can be transferred to two qubits by means of bi-local unitaries involving a mode and a qubit. Such a strategy has been used in Ref.~\cite{paternostro06}, where the entanglement in a mixed two-mode non-Gaussian state has been examined under the assumption that each mode-qubit subsystem interact via a local Jaynes-Cummings model. Although the technique is fundamentally interesting in light of the intricate relationship between systems of different dimensions, it is of not easy generalization to the multipartite scenario and we can at best provide only a lower bound to non-classical correlations, an exact quantification still being quite elusive. The lack of efficient ways to quantify entanglement motivates an assessment of non-classicality of ETS in terms of multipartite non-locality very appealing and, basically, the only rigorous way to ascertain whether or not a given non-Gaussian state of many modes is righteously entangled or not. This is the perspective which we will take in our study.

Although one may be tempted to identify one with the other, quantum non-locality and quantum entanglement in multipartite settings are concepts which should be approached carefully. In fact, it is straightforward to realize that the violation of an $n$-particle Bell's inequality by an $n$-particle entangled state is not sufficient to guarantee that genuine multipartite entanglement is shared by the system's elements. The non-local nature witnessed by the violation of a Bell's inequality might well be entailed simply by an entangled state involving only $m<n$ particles. In 1987, Svetlichny addressed this point by deriving an inequality, for the tripartite case, that is able to discriminate between two- and three-particle entanglement~\cite{svetl}. Such inequality is satisfied by restricted local realistic models allowing for a degree of two-particles non-locality. 

We now give a brief account of the inequality formulated by Svetlichny. Let us consider Mermin's version of Bell's inequality for three spin-$1/2$ particles~\cite{mermin}. In one of its possible formulations, {\it Mermin function} ${\cal M}$ is built from the correlation function ${\cal C}(abc)$ as 
\begin{equation}
\label{mermin1}
{\cal M}={\cal C}(abc_1)+{\cal C}(ab_1c)+{\cal C}(a_1bc)-{\cal C}(a_1b_1c_1),
\end{equation}
where the pairs $(k,k_1)$ with $k=a,b,c$ are the two dichotomic outcomes of an observable measured at the detection stage for one of the particles. Here, ${\cal C}(abc)$ is the statistical correlation function for measurements having outcomes $a,b$ and $c$ respectively. For local realistic theories (LRT), it is $|{\cal M}|\le{2}$. By exchanging $k$ with $k_1$ in Eq.~(\ref{mermin1}), one gets a new Mermin function ${\cal M}_1$ that, under LRT assumptions, obviously satisfies (in modulus) the same bound as ${\cal M}$. Therefore, by taking the {\it Svetlichny function} ${\cal S}({\bf a},{\bf b},{\bf c})=|{\cal M}+{\cal M}_1|$, we get
\begin{equation}
\label{svetequ}
\begin{aligned}
&{\cal S}({\bf a},{\bf b},{\bf c})=|{\cal C}(abc_1)+{\cal C}(ab_1c)+{\cal C}(a_1bc)+{\cal C}(abc)\\
&-{\cal C}(a_1b_1c)-{\cal C}(a_1bc_1)-{\cal C}(ab_1c_1)-{\cal C}(a_1b_1c_1)|\le{4},
\end{aligned}
\end{equation}
where ${\bf k}\!=\!(k,k_1)$. Quantum mechanics, on the other hand, predicts the existence of genuinely tripartite quantum correlated states violating such a bound. In particular, when the correlations ${\cal C}(abc)$ are evaluated over a tripartite GHZ state~\cite{ghz}, the value $4\sqrt{2}$ is obtained for the Svetlichny function, which is the maximum value achievable for any tripartite state. Svetlichny inequality (SI) is maximally violated by projecting each particle $j$ in a GHZ state onto the eigenstates $\ket{\pm}_j$ of the observable $\cos\vartheta\hat{\sigma}_x+\sin{\vartheta}\hat{\sigma}_y$~($\vartheta=\theta,\phi,\mu$), where $\hat{\sigma}_{x,y,z}$ are the three Pauli matrices. A straightforward calculation gives the correlation function 
\begin{equation}
{\cal C}(\theta,\phi,\mu)=\cos(\theta+\phi+\mu).
\end{equation}
 For $\theta=\theta_1+\pi/2=3\pi/4,\phi=-\mu_1=\pi/2$ and $\mu=\phi_1=0$, we have $|{\cal M}|=|{\cal M}_1|=2\sqrt{2}$ and $|{\cal S}|=4\sqrt{2}$, which shows violation of both Svetlichny and Mermin inequality. As discussed by Cereceda~\cite{cereceda}, SI is a righteous Bell inequality for the tripartite case and emerges as a valuable tool for the unambiguous ascertain of the existence of genuine tripartite entanglement and tripartite non-locality for any three-particle state. To this task, the use of a standard Mermin's inequality~\cite{mermin} is not sufficient: on the contrary, quantum correlations that violate SI are strong enough to maximally violate Mermin's inequality as well. The inequality by Svetlichny has been independently extended to the $n$-partite scenario in Refs.~\cite{S&S,collins}. Very recently, Lavoie {\it et al.} have experimentally demonstrated the violation of SI by a tripartite GHZ state~\cite{ghz} encoded in the polarization degrees of freedom of three photons in a linear optics setup~\cite{lavoie}.



\subsection{Generation of GHZ/W-like ETS and violation of Svetlichny inequality}
\label{gen}

\begin{figure}[hb]
\psfig{figure=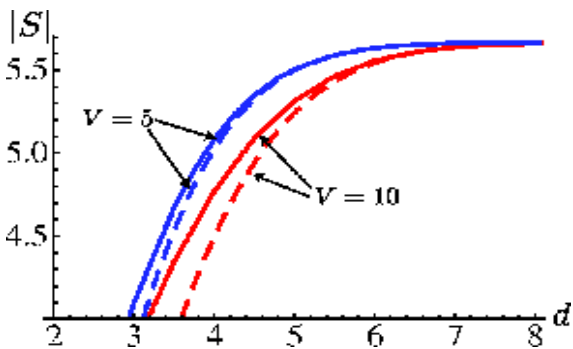,width=7.5cm,height=4.5cm}
\caption{Comparison between the Svetlichny function associated with state $\rho^{(3)}_{ghz,1}$ (full lines) and $\rho^{(3)}_{ghz,2}$ (dashed ones) for $V=5,10$ and $\eta=0.3$ (assumed to be the same for every detector).}
\label{alternativo}
\end{figure}

We now address tripartite non-locality of ETS. In our proposal, three local parties are each provided with one mode $j=1,2,3$ of a  tripartite ETS that has been off-line prepared using single-mode displaced thermal states~\cite{jr06,jacobmaurotim}.
These are defined as
$\rho^{\rm th}_j(V,d)=\int d^2\alpha P_\alpha^{\rm th}(V,d)
|\alpha\rangle_j\langle\alpha|$, where $P_\alpha^{\rm th}(V,d)={2}[\pi(V-1)]^{-1}
e^{-\frac{2|\alpha-d|^2}{V-1}}$ is a Gaussian function with center $d$ (with respect to the origin of the phase space) and variance proportional to $V=2\nbar+1$, where $\nbar$ is the mean thermal occupation number of the mode. Here, $\ket{\alpha}_j$ is a coherent state of mode $j$, which has associated  bosonic operators $\hat{a}_j$ and $\hat{a}^\dag_j$. Displaced thermal states are the building blocks for the construction of a tripartite GHZ-like ETS state, as described in Figs.~\ref{generazione} {\bf (a)} and {\bf (b)}. The scheme in panel {\bf (a)} is probabilistic and based on conditional displacements of each of the three modes upon interaction with a two-level ancilla $A$ having logical states $\ket{0}_A$ and $\ket{1}_A$. This is realized by enforcing the mode-ancilla coupling $\hat{H}_{Aj}=\hbar\Omega|1\rangle_A\langle{1}|\hat{a}^\dag_j\hat{a}_j$ and upon preparation of $A$ in $\ket{+}_A=(\ket{0}_A+\ket{1}_A)/\sqrt 2$. Nonlinear media with free-traveling optical fields~\cite{free1,free2}
or dispersive interactions within optical/microwave cavities~\cite{cavities}
may be used to implement the required interactions~\cite{jr06,paternostro06}. The state of $A$ is eventually projected onto the state basis $\{\ket{+}_A,\ket{-}_A\}$ (with ${}_{A}\!\langle-|+\rangle_A=0$), as shown in Fig.~\ref{generazione} {\bf (a)}. The scheme in Fig.~\ref{generazione} {\bf (b)}, on the other hand, relies on interference at appropriately-arranged beam splitters, in a way completely analogous to the proposal put forward in Ref.~\cite{jeongNguyen}. Having off-line prepared state $\rho^{\text{th}'}(V,d)={\cal N}_+\int{d}^2\alpha P^{\text{th}}_\alpha(V,d)\ket{s^+_\alpha}\bra{s^+_\alpha}$ of mode $1$ with
\begin{equation}
\label{deff}
\ket{s^\pm_\alpha}\propto(\ket{\alpha}\pm\ket{-\alpha}),
\end{equation}
a three-mode GHZ-like ETS is achieved by superimposing it to two additional modes in their vacuum state. State $\rho^{\text{th}'}(V,d)$ can be prepared similarly to what is shown in panel {\bf (a)}, i.e. by letting a displaced thermal state of mode $1$ interact, according to $\hat{H}_{A1}$ and for a time $\pi/\Omega$, with a two-level ancilla prepared in $\ket{+}_A$. The latter is eventually projected onto the basis $\{\ket{+}_A,\ket{-}_A\}$ to leave mode $1$ in  $\rho^{\text{th}'}(V,d)$~\cite{jr06}. The scheme in Fig.~\ref{generazione} {\bf (b)} creates the state
\begin{equation}
\label{uno}
\rho^{(3)}_{ghz,1}{=}{\cal N}_1{\int d^2}\alpha{P}^{\text{th}}_\alpha(V,d)\ket{\text{GHZ}(\tilde\alpha,\tilde\alpha,\tilde\alpha)}_{123}\!\bra{\text{GHZ}(\tilde\alpha,\tilde\alpha,\tilde\alpha)}
\end{equation}
with $\tilde\alpha=\alpha/\sqrt{3}$, while the conditional approach in Fig.~\ref{generazione} {\bf (a)} gives
\begin{equation}
\label{due}
\begin{aligned}
\rho^{(3)}_{ghz,2}={\cal N}_2\int&{d}^2\alpha\,{d}^2\beta\,{d}^2\zeta{P}^{\text{th}}_\alpha(V,d){P}^{\text{th}}_\beta(V,d){P}^{\text{th}}_\zeta(V,d)\\
&\times\ket{\text{GHZ}(\alpha,\beta,\zeta)}_{123}\!\bra{\text{GHZ}(\alpha,\beta,\zeta)}
\end{aligned}
\end{equation}
with ${\cal N}_{1,2}$ being normalization factors and ${\alpha,\beta,\zeta\in{\mathbb C}}$. In these equations we have used the (unnormalized) GHZ-like entangled coherent state 
\begin{equation}
\label{ghz}
\ket{\text{GHZ}(\alpha,\beta,\zeta)}_{123}=(\hat\openone+\otimes^3_{j=1}e^{i\pi\hat{a}^\dag_j\hat{a}_j})\ket{\alpha,\beta,\zeta}_{123}
\end{equation}
where $\openone$ is the identity operator.  Although conceptually analogous, the cases encompassed by Eq.~(\ref{uno}) and (\ref{due}) are technically different and the class of states represented by $\rho^{(3)}_{ghz,2}$ turns out to be analytically easier to treat. Therefore, in order to provide a clear interpretation of our results, in what follows we study the case of Eq.~(\ref{due}) in detail, leaving the assessment of Eq.~(\ref{uno}) to the numerical results shown in Fig.~\ref{alternativo}. It is worth noting that both the tripartite GHZ- and W-like entangled coherent states have been shown to violate Mermin's inequality~\cite{jeongNguyen}.

Our task here is to give a clear-cut account of the main results of our investigation, providing at the same time an intuition of the physical features behind the observation of non-classical multipartite correlations in the classes of multiparty ETS-based states addressed here. We aim at showing violation of SI under unfavorable conditions such as high-temperature local bosons and ``fuzzy"  measurements.
We address the case of a set of local rotations performed over each of the modes entering a GHZ-like ETS state. Explicitly, we consider the set of angles $(\theta,\gamma)$ and the associated transformation  
\begin{equation}
\hat{R}_1(\theta,\gamma)=\left(\begin{array}{cc}\sin(\theta/2)&e^{i\gamma}\cos(\theta/2)\\
e^{-i\gamma}\cos(\theta/2)&-\sin(\theta/2)\end{array}\right)
\end{equation}
to be applied to the vector of coherent states $\begin{pmatrix}\ket{\alpha}_1&\ket{-\alpha}_1\end{pmatrix}^T$ for mode $1$. Modes $2$ and $3$ experience similar transformations, each determined by the pairs of angles $(\phi,\delta)$ and $(\mu,\nu)$ respectively. An implementation of such a set of rotations using non-linear interactions and phase-space displacements is discussed in Refs.~\cite{stobinska,jacobmaurotim} Following the arguments by Svetlichny~\cite{svetl} illustrated above, we restrict the set of local rotations over the modes at hand to $\theta=\phi=\mu=\pi/2$ and, unless stated otherwise, indicate the rotation operators as $\hat{R}_{j}(\sigma)$ ($\sigma=\gamma,\delta,\nu$). The projections needed in order to evaluate joint probabilities and correlation functions as described in Eq.~(\ref{svetequ}) are instead implemented by dichotomizing the outcomes of homodyne measurements performed over the three modes in $\rho^{(3)}_{ghz,2}$. We associate a logical $+1$ ($-1$) to a positive (negative) homodyne signal, as illustrated in Fig.~\ref{generazione} {\bf (c)}. For the tripartite case, we calculate the correlation function as 
\begin{equation}
C(\gamma,\delta,\nu)=P_{+++} -P_{---} +\Pi_{}({P}_{+--}-{P}_{++-}),
\end{equation}
where $\Pi$ represents a permutation of the pedices appearing in the conditional probabilities $P_{klp}$ (with $k,l,p=\pm$). The latter are easily calculated as
\begin{equation}
P_{klp}=\!\int_{{\cal R}_k}{d}{x}\!\int_{{\cal R}_l}{d}{y}\!\int_{{\cal R}_p}{d}{z}~\langle{x},y,z|\rho^{(3)'}_{ghz,2}|x,y,z\rangle,
\end{equation}
where we have introduced the rotated GHZ-like ETS ${\rho^{(3)'}_{ghz,2}=\hat{R}_1(\gamma)\hat{R}_2(\delta)\hat{R}_3(\nu)\rho^{(3)}_{ghz,2}\hat{R}^\dag_1(\gamma)\hat{R}^\dag_2(\delta)\hat{R}^\dag_3(\nu)}$ and the positive (negative) domain of integration ${\cal R}_{+}=[0,\infty)$ (${\cal R}_{-}=(-\infty,0]$). It is thus found that
\begin{equation}
\label{correlationGHZ}
C_{ghz,2}(\gamma,\delta,\nu)={\cal N}_{ghz,2}(\gamma,\delta,\nu)\cos(\gamma+\delta+\nu){ \mbox{Erf}}^3[\frac{\sqrt{2}d}{\sqrt{V}}]
\end{equation}
with $\text{Erf}[x]$ the error function of argument $x$.  The appearance of the error function in the result above is noteworthy. It arises from the finite-range integration over the products of the Gaussian $P^{\text{th}}_{\alpha}$ function entering the definition of a thermal state and the one resulting from the homodyne probing of the fields' quadratures. When ${d\gg{\sqrt{V/2}}}$, ${ \mbox{Erf}}[\sqrt{2}d/\sqrt{V}]\rightarrow{1}$ with ${\cal N}_{ghz,2}\rightarrow{1}$, making the correlation function identical to the one obtained for the case of a tripartite spin-$1/2$ GHZ state~\cite{lavoie}. This is the key of our results. As confirmed by the analysis performed in the remainder of the paper, we can generalize the functional form in Eq.~(\ref{correlationGHZ}) and claim that the correlation function for any ETS-based multipartite state will encompass an angular part that, under ideal rotations, is identical to the one achieved for spin-$1/2$ particles. The modulus of the correlation function, though, also depends in a crucial way on the degree of distinguishability of the multipartite-state components, as encompassed by the error function. 
This accounts for the clear {\it universal} behaviour of the correlation functions associated with ETS-based states against the key parameter of the local thermal components. 

\begin{figure}[t]
\psfig{figure=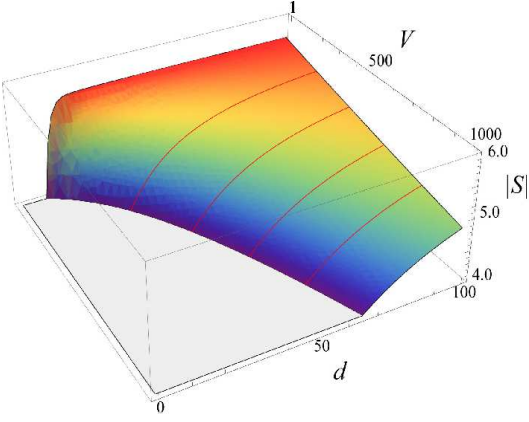,width=7cm,height=5.5cm}
\caption{Violation of Svetlichny inequality by a tripartite GHZ-like ETS under effective rotations and inefficient homodyne detection (efficiency $\eta=0.1$). We show the Svetlichny function against displacement $d$ and variance $V$ of the local thermal distributions. The floor of the plot is given by the local realistic bound of $4$. For $d\gg{\sqrt{V}}$, the upper bound of $4\sqrt{2}$ is achieved, exactly as it would be for a pure GHZ state of three spin-$1/2$ particles under sharp measurements.}
\label{ineff}
\end{figure}

When inefficient homodyne detectors are considered, all having the same detection efficiency $\eta$, the correlation function is easily found from $C_{ghz,2}(\gamma,\delta,\nu)$ with the replacements $d\rightarrow\eta{d}$ and $V\rightarrow{1}+\eta^2(V-1)$ in the error function appearing in Eq.~(\ref{correlationGHZ}). So, the effect of an inefficient detector is just to slow down the saturation of the $d$-dependent term in the correlation function. This simply means that, for  set values of $V$ and for $\eta<1$, a larger value of $d$ is required to achieve the maximum allowed degree of violation of Svetlichny inequality. More in detail, for $V\gg{1}$ we need to displace the local thermal states by $d_{\eta}\simeq{d}{\sqrt{1+(V\eta^2)^{-1}}}$ in order for the corresponding correlation function to reach the value of $C_{ghz,2}(\gamma,\delta,\nu)$ corresponding to $\eta=1$ (i.e. for perfectly efficient detectors). The scaling is quite favourable at large values of $V$, which shows that highly thermal states that are initially considerably displaced with respect to the origin of phase space are extremely insensitive to the effects of inefficient homodyne detection. One could even ignore the actual detection inefficiency and, at large $V$, displace the local states as if the homodyners were ideal, without affecting the performance of the scheme. From the above discussion it is straightforward to understand that, under the present conditions and for the GHZ-like ETS state addressed above, we get
\begin{equation}
\label{sveteffect}
S^{ghz,2}_\eta({\bm \gamma},{\bm \delta},{\bm \nu})\propto{\cal S}^{ghz,2}({\bm \gamma},{\bm \delta},{\bm \nu}){ \mbox{Erf}}^3[\frac{\sqrt{2}d\eta}{\sqrt{1+\eta^2(V-1)}}]
\end{equation}
where ${\cal S}^{ghz,2}({\bm \gamma},{\bm \delta},{\bm \nu})$ is the Svetlichny function for the ideal case of spin-$1/2$ particles and the proportionality sign is due to a factor that tends to $1$ as $d^2\gg(\eta^{-2}+V-1)/2$. This shows that {\sl non-classicality as witnessed by the violation of Svetlichny inequality will be observed up to its maximum allowed degree even for large initial temperatures of the multipartite state}. Fig.~\ref{ineff} shows the Svetlichny function against $V$ and $d$ for $\eta=0.1$. Maximum violations of the inequality can be observed for any value of $V$ by choosing a sufficiently large $d$.  The parameter $d$  is the ``knob" to tune in order to optimize the non-classical properties of a given state at an assigned value of the thermal spread $V$. Intuitively this means that the entanglement in an ETS is a delicate trade-off between the distinguishability of its state components, as measured by their mutual distance $d$ in phase-space, and the width $V$ of each thermal distribution. When the Gaussian probability functions defining each thermal state are so large that they overlap significantly, the state components become quasi-indistinguishable and entanglement is correspondingly destroyed. It should thus be clear that, per assigned value of $V$, a way to counteract such Òentanglement washing-out effectÓ is to make the
state components sufficiently distinguishable in phase space, which implies the increase of $d$.

An analogous behavior is found also for the alternative form of GHZ-like ETS $\rho^{(3)}_{ghz,1}$. In this case, though, due to the complications related to the calculations that are necessary to get the correlation functions, an analytic expression for the Svetlichny function turns out to be not possible. Nevertheless, the inequality can be well assessed, quantitatively, by relying on a numerical approach that allows us to assess the behavior of $S^{ghz,1}_\eta({\bm \gamma},{\bm \delta},{\bm \nu})$ against $d$, at set values of $V$ and $\eta$. While, qualitatively, the very same features highlighted for the case of $\rho^{(3)}_{ghz,2}$ hold also in this case (monotonic growth of the Svetlichny function with $d$ and saturation to the degree of maximum violation for $d\gg\sqrt V$), it is interesting to determine if one of the state-generation strategies proposed here  offers any advantage when evaluated against the violation of Svetlichny inequality. In Fig.~\ref{alternativo} we show a quantitative comparison between $S^{ghz,1}_\eta({\bm \gamma},{\bm \delta},{\bm \nu})$ and $S^{ghz,2}_\eta({\bm \gamma},{\bm \delta},{\bm \nu})$ for two values of $V$ (arbitrary choices) and using the same detection inefficiency $\eta$.  We can clearly see that, despite the evident similarities in the form of the two Svetlichny functions, the generation protocol illustrated in Fig.~\ref{generazione} {\bf (b)} turns out to be slightly more convenient: the local realistic bound of 4 [see Eq.~(\ref{svetequ})] is surpassed for slightly smaller values of $d$, when the beam splitter-based generation scheme is used. Such a trend is shown regardless of the value  of $V>1$ and $\eta$ chosen for our numerical test and we have strong numerical evidence that the distance between the Svetlichny functions associated with the two generation schemes opens up (at small values of $d$) as $V$ increases. In Sec.~\ref{quadripartite} we will see that a similar result holds for a quadripartite linear cluster state. It is worth stressing the existence of another possibility to generate tripartite GHZ-like ETS by means of the scheme in Fig.~\ref{generazione} {\bf (b)}. Instead of relying on a resource having the form of  $\rho^{\text{th}'}(V,d)$, one could well use state $\rho^{\text{th}''}(V,d)=\hat{U}\rho^{\text{th}}(V,d)\hat{U}^{-1}$, where the unitary evolution $\hat{U}=e^{-i\pi(\hat{a}^\dag\hat{a})^2/2}$ is obtained by means of a self-Kerr non-linear medium. In this case, a single-mode resource having the form $\rho^{\text{th}''}(V,d)\propto\int{d}^2\alpha P^{\text{th}}_\alpha(V,d)|r^+_\alpha\rangle\langle{r^+_\alpha}|$ is achieved, where $\ket{r^+_\alpha}\propto(\ket{\alpha}+i\ket{-\alpha})$. We have checked that, for the GHZ-like ETS that is generated by using $\rho^{\text{th}''}(V,d)$ in the protocol of Fig.~\ref{generazione} {\bf (b)}, there is always a set of local rotations that, complemented by dichotomized homodyne measurements, allow for the violation of the tripartite Svetlichny inequality in a way fully analogous to what has been addressed so far.

\begin{figure}[h]
\psfig{figure=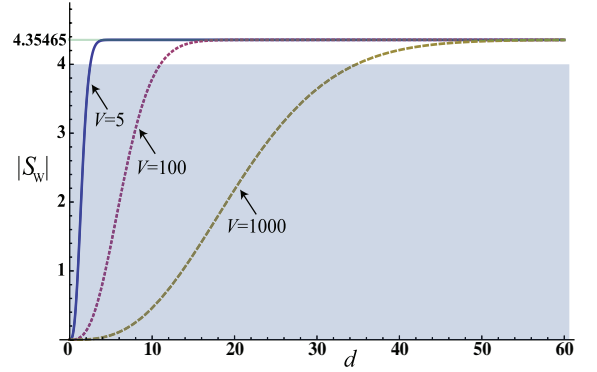,width=7.5cm,height=5cm}
\caption{Svetlichny function for a tripartite W-like ETS at various values of $V$. The local realistic bound is $4$ (shaded region of the plot). The maximum achieved violation of Svetlichny inequality, reached at large values of $d$, is quantitatively the same as in the spin-$1/2$ case.}
\label{wstate}
\end{figure}

The qualitative features discussed above do not depend on the form of the multipartite state being considered. In fact, by following arguments similar to those valid for a GHZ-like ETS state, one can easily verify that the correlation function corresponding to an ETS version of the W state, which can be generated as discussed in~\cite{jeongNguyen}, reads (for $\eta=1$) 
\begin{equation}
\label{correlationW}
C_w(\theta,\phi,\mu)\!\propto\![\cos\theta\cos\phi\cos\mu+2\cos(\theta+\phi+\mu)]{ \mbox{Erf}}^3[\frac{\sqrt{2}d}{\sqrt{V}}].
\end{equation}
The angular part is identical to what has been obtained by Cereceda~\cite{cereceda} for a tripartite spin-$1/2$ W state $(1/\sqrt{3})\sum^3_{j=1}\hat{\sigma}_{x,j}\ket{000}_{123}$. This class of states is known to violate Svetlichny inequality by $\simeq0.355$~\cite{cereceda}, which is achieved by projecting each party onto the eigenstates of $\cos\vartheta\hat{\sigma}_z+\sin\vartheta\hat{\sigma}_x~(\vartheta=\theta,\phi,\mu)$. In our formalism, this is equivalent to applying the local rotations $\hat{R}_{j}(2\arctan[(\cos\vartheta-1)/\sin\vartheta],0)$ to each mode and perform dichotomic homodyne measurements as described above. The corresponding Svetlichny function $S_{W}$ is shown in Fig.~\ref{wstate} for several values of $V$ and $\eta=1$. Evidently, a large enough ratio $d/\sqrt{V}$ makes a W-like ETS state violate the tripartite Svetlichny inequality up to the maximum degree allowed to the spin-$1/2$ counterpart of such states. The inclusion of detection inefficiency and the corresponding results are perfectly analogous to what was discussed in this Section regarding the GHZ-like ETS case. 

\section{QUADRIPARTITE case: GHZ AND CLUSTER-like ETS}
\label{quadripartite}

\subsection{Quadripartite GHZ-like ETS}

We now extend our study to the case of quadripartite ETS. As  a first, simple step, we address the extension of our previous study to non-locality of a quadripartite GHZ-like ETS. As done in the tripartite case, we focus our attention to the general state-engineering scheme depicted in Fig.~\ref{generazione} {\bf (a)}, which can be straightforwardly generalized to the four-partite scenario by considering a fourth local displaced thermal state and, correspondingly, an additional ancilla-mode interaction. For equal local temperatures and displacements, the resulting quadripartite GHZ-like ETS takes the form
\begin{equation}
\label{quadghz}
\begin{aligned}
\rho^{(4)}_{ghz}&={\cal N}_3\int{d}^2\alpha\,{d}^2\beta\,{d}^2\gamma\,{d}^2\delta{P}^{\text{th}}_\alpha(V,d){P}^{\text{th}}_\beta(V,d){P}^{\text{th}}_\gamma(V,d)\\
&\times{P}^{\text{th}}_\delta(V,d)\ket{\text{GHZ}(\alpha,\beta,\gamma,\delta)}_{1234}\bra{\text{GHZ}(\alpha,\beta,\gamma,\delta)}
\end{aligned}
\end{equation}
\begin{figure}[hb]
\psfig{figure=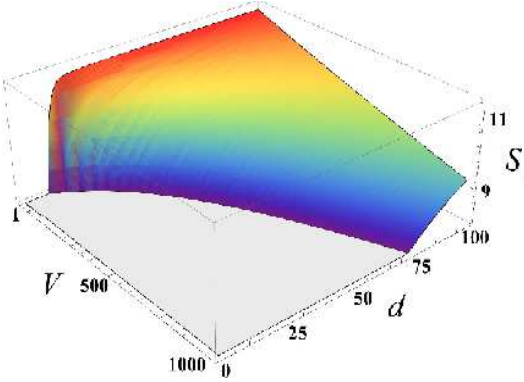,width=7cm,height=5.0cm}
\caption{Svetlichny function for a four-partite GHZ-like ETS. As before, we have used $\eta=0.1$. Local realisitic theories enforce a bound of  $8$ while the maximum achievable quantum mechanically is 8$\sqrt{2}$.}
\label{quadstatesViolation}
\end{figure}
with ${\cal N}_3$ a normalisation factor, ${\alpha,\beta,\gamma,\delta\in{\mathbb C}}$ and $\ket{\text{GHZ}(\alpha,\beta,\delta,\gamma)}{=}\ket{\alpha,\beta,\gamma,\delta}+\ket{-\alpha,-\beta,-\gamma,-\delta}$. In Ref.~\cite{S&S}, Seevinck and Svetlichny have generalized the original argument by Svetlichny and provided the explicit form for the inequality to assess in the quadripartite scenario. This reads
\begin{equation}
\label{svetequ2}
\begin{aligned}
&{\cal S}({\bf a},{\bf b},{\bf c},{\bf d}){=}|{{\cal C}(abcd)-{\cal C}(a_1bcd)-{\cal C}(ab_1cd)-{\cal C}(abc_1d)}\\
&{+{\cal C}(a_1b_1cd_1)+{\cal C}(a_1bc_1d_1)+{\cal C}(ab_1c_1d_1)+{\cal C}(a_1b_1c_1d_1)}\\
&{-{\cal C}(abcd_1)-{\cal C}(a_1b_1cd)-{\cal C}(a_1bc_1d)-{\cal C}(a_1bcd_1)}\\
&{-{\cal C}(ab_1c_1d)-{\cal C}(ab_1cd_1)-{\cal C}(abc_1d_1)+{\cal C}(a_1b_1c_1d)}|\le{8}.
\end{aligned}
\end{equation}
Quantum mechanically, there are genuinely multipartite entangled states that violate the local realistic bound of 8. In particular, a four-party GHZ state violates this inequality maximally, achieving $|{\cal S}({\bf a},{\bf b},{\bf c},{\bf d})|=8\sqrt 2$ for the same set of local operations considered in Sec.~\ref{tripartite}. When tested using the state in Eq.~(\ref{quadghz}), the four-party Svetlichny function shows a behavior similar to what is shown in Fig.~\ref{ineff}: for any specified value of $V$ and of the detection efficiency $\eta$, a sufficiently large displacement ensures the violation of SI up to the maximum allowed value $8\sqrt{2}$. Fig.~\ref{quadstatesViolation} shows a significant example of such a behavior. 

\subsection{Quadripartite linear cluster-like ETS}

\begin{figure*}[ht]
\center{\psfig{figure=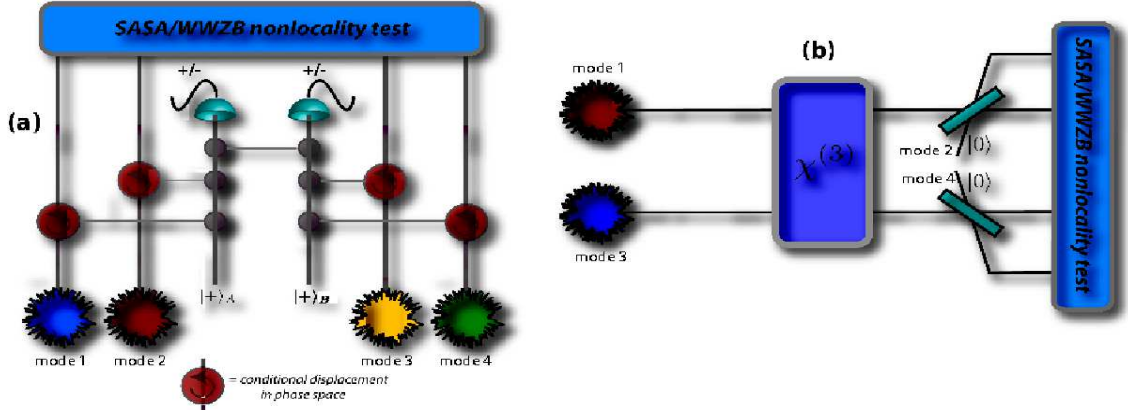,width=15cm,height=5.5cm}}
\caption{{\bf (a)} Conditional scheme for the generation of a four-partite linear cluster state via local thermal states, conditional displacements and postselection. The interpretation of the symbols is the same as in the caption of Fig.~\ref{generazione}. Differently from the GHZ-like ETS case, here we need two ancillary two-level systems and their mutual interaction. In panel {\bf (b)} we show an alternative way of generating a linear cluster-ETS (see Ba an and Hoa in Ref.~\cite{others}). Two thermal states of annihilation operators $\hat{a}_{j}$ ($j=1,2$) enter a non-linear cross-Kerr medium where the interaction $\hat{H}_{\chi^{(3)}}=\hbar\chi\hat{a}^\dag_1\hat{a}_1\hat{a}^\dag_2\hat{a}_2$ takes place for a time $\pi/\chi$. The output modes are then superimposed to two ancillary modes prepared in their vacuum state at $50:50$ beam spitters. Local phase shifters (not shown) complete the generation process.
}
\label{generazione2}
\end{figure*}

Unlike the tripartite case, where only two inequivalent classes of multipartite entangled states exist~\cite{cirac}, a plethora of choices are available in the quadripartite scenario. Among the states that can be considered, the class of cluster states is dressed of considerable interest in virtue of the role they play in the one-way paradigm for quantum computing~\cite{briegel}. It is thus interesting and relevant to assess the case embodied by such states and to build up their ETS counterpart. In what follows, we focus our attention to the analogue of the spin-$1/2$ cluster state 
\begin{equation}
\ket{C_{1/2}}=(\ket{0000}+\ket{1100}+\ket{0011}-\ket{1111})/2
\end{equation}
with $\{\ket{0},\ket{1}\}$ the logical states of each particle and the subscript $1/2$ reminding us that this form holds for a register of spins.  Schemes for the generation of coherent state-encoded linear cluster states having a structure similar to the above have been put forward~\cite{fernando} and their usefulness as quantum channels for teleportation has been assessed by one of us~\cite{fernando2}. Here we describe suitable generation protocols for linear cluster-like ETS and then use them to test genuine quadripartite non-locality. The first scheme we propose is a straightforward modification of both the protocol put forward in Ref.~\cite{fernando} and the general procedure utilized in order to generate GHZ-like ETS [cfr. Fig.~\ref{generazione} {\bf (a)}].  It is based on conditional displacement operations actuated by two ancillary two-level systems resulting from the same sort of interaction Hamiltonian introduced in Sec.~\ref{tripartite}. Differently from the scheme for GHZ-like ETS, here a mutual coupling between the ancillary systems has to be arranged according to $\hat{H}_{12}=\hbar\Omega|11\rangle_{AB}\langle{11}|$, where $A$ and $B$ are labels identifying the two ancillae, whose state after the network of couplings depicted in Fig.~\ref{generazione2} {\bf (a)} is projected onto the $\{\ket{+},\ket{-}\}_{A,B}$ bases. Upon detection of $\ket{++}_{AB}$, the state of modes $j=1,2,3$ and 4 becomes
\begin{equation}
\label{clus}
\begin{aligned}
&\rho^{(1)}_{cluster}=\int{d}^2\alpha\,{d}^2\beta\,{d}^2\gamma\,{d}^2\delta{P}^{\text{th}}_\alpha(V,d){P}^{\text{th}}_\beta(V,d)\\
&\times{P}^{\text{th}}_\gamma(V,d){P}^{\text{th}}_\delta(V,d)\ket{\text{C}(\alpha,\beta,\gamma,\delta)}_{1234}\bra{\text{C}(\alpha,\beta,\gamma,\delta)}
\end{aligned}
\end{equation}
with ${\alpha,\beta,\gamma,\delta\in{\mathbb C}}$. By identifying a logical $\ket{0}$ ($\ket{1}$) state of mode $1$ with a coherent state of amplitude $\ket{\alpha}_1$ ($\ket{-\alpha}_1$) and using an analogous encoding for the remaining modes, state $\ket{\text{C}(\alpha,\beta,\gamma,\delta)}$ turns out to be the coherent state-encoded linear cluster state 
\begin{equation}
\begin{aligned}
&\ket{\text{C}(\alpha,\beta,\gamma,\delta)}=\frac{1}{2}(\ket{\alpha,\beta,\gamma,\delta}+\ket{\alpha,\beta,-\gamma,-\delta}\\
&+\ket{-\alpha,-\beta,\gamma,\delta} -\ket{-\alpha,-\beta,-\gamma,-\delta}).
\end{aligned}
\end{equation}    
We will address an alternative state-engineering procedure in the next Subsection. The non-locality properties of spin-$1/2$ cluster states have been addressed in a seminal work by Scarani, Acin, Schenck and Aspelmeyer~\cite{SASA}, which we dub from now on as SASA. Motivated by the fact that a linear cluster state does not maximally violate a four-partite Mermin inequality, SASA looked for an inequality that is maximally violated by quadripartite linear cluster states. By using the  stabilizer operators, they have formulated a simple four-term correlator, here named SASA function. More in detail, for the form of the linear cluster state under scrutiny here, the spin-$1/2$ SASA function is obtained as the expectation value of the correlation operator
\begin{equation}
\label{SASA}
\begin{aligned}
&\hat{O}_{\text{SASA}}{=}\hat{\sigma}_{z,1}\otimes\hat\openone_{2}\otimes\hat{\sigma}_{x,3}\otimes\hat{\sigma}_{z,4}-\hat{\sigma}_{z,1}\otimes\hat\openone_{2}\otimes\hat{\sigma}_{y,3}\otimes\hat{\sigma}_{y,4}\\
&+\hat{\sigma}_{x,1}\otimes\hat{\sigma}_{y,2}\otimes\hat{\sigma}_{y,3}\otimes\hat{\sigma}_{x,4}+\hat{\sigma}_{x,1}\otimes\hat{\sigma}_{y,2}\otimes\hat{\sigma}_{x,3}\otimes\hat{\sigma}_{y,4}.
\end{aligned}
\end{equation} 
calculated over the cluster state. Classically, by associating $\pm{1}$ to any of the operators involved in $\hat{O}_{\text{SASA}}$, it is straightforward to check that the inequality ${|\langle\hat{O}_{\text{SASA}}\rangle|\le{2}}$ holds. Quantum mechanically, by construction, we get $\bra{C_{1/2}}\hat{O}_{\text{SASA}}\ket{C_{1/2}}=4$. No other state can achieve a larger value, thus making the SASA inequality optimal for the class of states under scrutiny. Remarkably, such inequality has been experimentally tested using the polarization degree of freedom of four photonic modes~\cite{exp}.  

\begin{figure}[b]
\psfig{figure=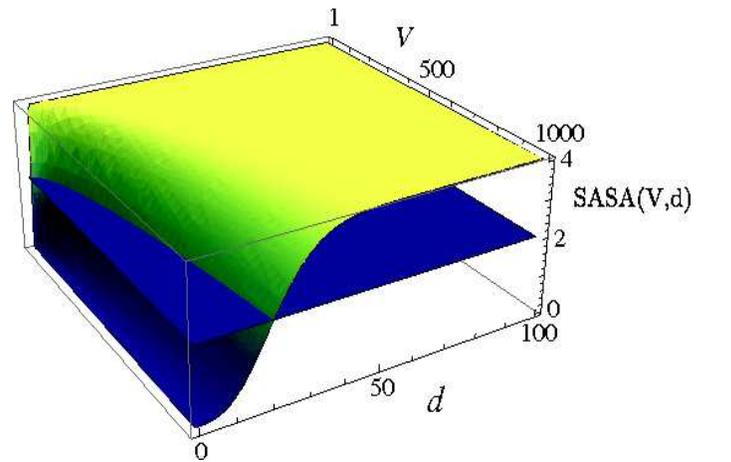,width=9cm,height=6.0cm}
\caption{Violation of SASA inequality by a quadripartite cluster-like ETS under effective rotations and homodyne detection. The SASA function is plotted against $V$ and $d$. The local realistic bound is shown by the plane at $\text{SASA}=2$.}
\label{C1}
\end{figure}
Our aim now is to prove that state $\rho_{cluster}$ defined in Eq.~(\ref{clus}) violates SASA inequality up to its maximum value when the local thermal states are sufficiently displaced in phase space with respect to the variance $V$. As done before, the necessary correlators are implemented by dichotomising the outcomes of homodyne measurements performed over the four modes $j=1,2,3,4$ and using appropriate local operations $\hat{R}_j$'s. By following the lines depicted in Sec.~\ref{tripartite}, we get the SASA function
\begin{equation}
\label{sasafunc}
\text{SASA}(V,d)=2 \mbox{Erf}^3\!\!\left[\f{\sqrt{2}d}{\sqrt{V}}\right]\left(1+\mbox{Erf}\left[\f{\sqrt{2}d}{\sqrt{V}}\right]\right)
\end{equation}
where perfectly efficient homodyne detectors have been assumed, for the sake of clarity. Our analysis can be straightforwardly extended to the case of inefficient detectors simply by performing the replacements for $d$ and $V$ discussed for the case of a tripartite GHZ -like ETS state. Fig.~\ref{C1} shows the trend of Eq.~(\ref{sasafunc}) against $V$ and $d$ for a considerable range of values. For $V\sim10^3$, only moderate values of  $d$ are sufficient to violate the local realistic bound, although maximum violation occurs only at $d\gtrsim{50}$. Hence our linear cluster ETS maximally violates this inequality for any value of $V$ by choosing $d$ sufficiently large. 

\begin{figure}[t]
\psfig{figure=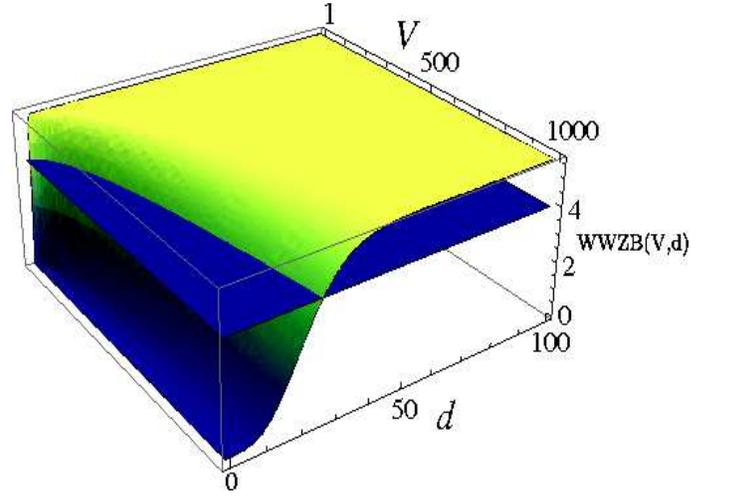,width=9cm,height=6.5cm}
\caption{Violation of WWZB inequality by a four party linear cluster-like ETS under effective rotations and homodyne detection. The local realistic bound is 4 and is indicated by the horizontal plane. For any $V$, a correspondingly large value of $d$ guarantees a violation of the WWZB inequality by a factor $\sqrt2$, in agreement with the spin-$1/2$ case.}
\label{C2}
\end{figure}

The fact that one of the parties in SASA has to use only a single measurement setting (the local observable for mode $2$ has to be $\hat{\openone}$ in two cases out of four) marks the fundamental difference between such an inequality and any Mermin/Svetlichny-like argument, where at least two measurement settings are required for each of the parties involved. Indeed, SASA inequality does not fall into the class of generalised Mermin inequalities, such as those proposed by Collins et al.~\cite{collins}, even more given that it stems from the use of stabilizer operators for the cluster state at hand. It is therefore interesting to test which is the behavior of our four-mode cluster-like ETS against Mermin-like inequalities. For this task, we consider the works by Werner-Wolf~\cite{WW} and \.Zukowski-Brukner~\cite{ZB}, who independently proposed equivalent and general inequalities for an arbitrary number of particles, using two measurement settings per party. Here we concentrate on the formulation by \.Zukowski and Brukner, whose inequality for spin-$1/2$ particles takes the form
\begin{equation}
\label{wwzbspin}
\begin{aligned}
&|\text{WWZB}|{=}|{\cal C}(abcd){+}{\cal C}(abcd_1){+}{\cal C}(abc_1d)-{\cal C}(abc_1d_1)\\
&+{\cal C}(ab_1cd){-}{\cal C}(ab_1cd_1){-}{\cal C}(ab_1c_1d){-}{\cal C}(ab_1c_1d_1)\\
&+{\cal C}(a_1bcd){-}{\cal C}(a_1bcd_1){-}{\cal C}(a_1bc_1d){-}{\cal C}(a_1bc_1d_1)\\&
{-}{\cal C}(a_1b_1cd){-}{\cal C}(a_1b_1cd_1){-}{\cal C}(a_1b_1c_1d){+}{\cal C}(a_1b_1c_1d_1)|\le4 
\end{aligned}
\end{equation}
with WWZB indicating the Werner-Wolf-\.Zukowski-Brukner parameter.  By using the spin-$1/2$ linear cluster state $\ket{C_{1/2}}$, it is seen that the local observable for particle $j$ needed to build up the correlations all have the form 
\begin{equation}
\hat{R}(\phi_j)= \f{1}{\sqrt{2}}\left(\begin{array}{lr} 1 & e^{i\phi_j}\\e^{-i\phi_j} & -1\end{array}\right),
\end{equation}
so that ${\bf k}=(\phi_1,\phi'_1)$ and analogous definitions for the remaining set of parameters. Numerical maximization of the left-hand side of Eq.~(\ref{wwzbspin}) over such sets leads to the violation of WWZB inequality by a factor $\sqrt 2$ for the set of angles  
$\phi_j=3\pi/16,\phi'_j=11\pi/16,~\forall{j}$. We retain such values for our test of non-locality involving the cluster-like ETS and calculate the appropriate correlators by means of our dichotomized homodyne measurements. The corresponding WWZB function is too lengthy to be given here and we thus rely on Fig.~\ref{C2} for a comprehensive account of our results, which show that violation up to the maximum allowed to the class of states addressed here is achievable under the same general working conditions valid in the case of the SASA inequality. 
        
\subsection{Alternative generation scheme for linear cluster-like ETS}

\begin{figure}[b]
\psfig{figure=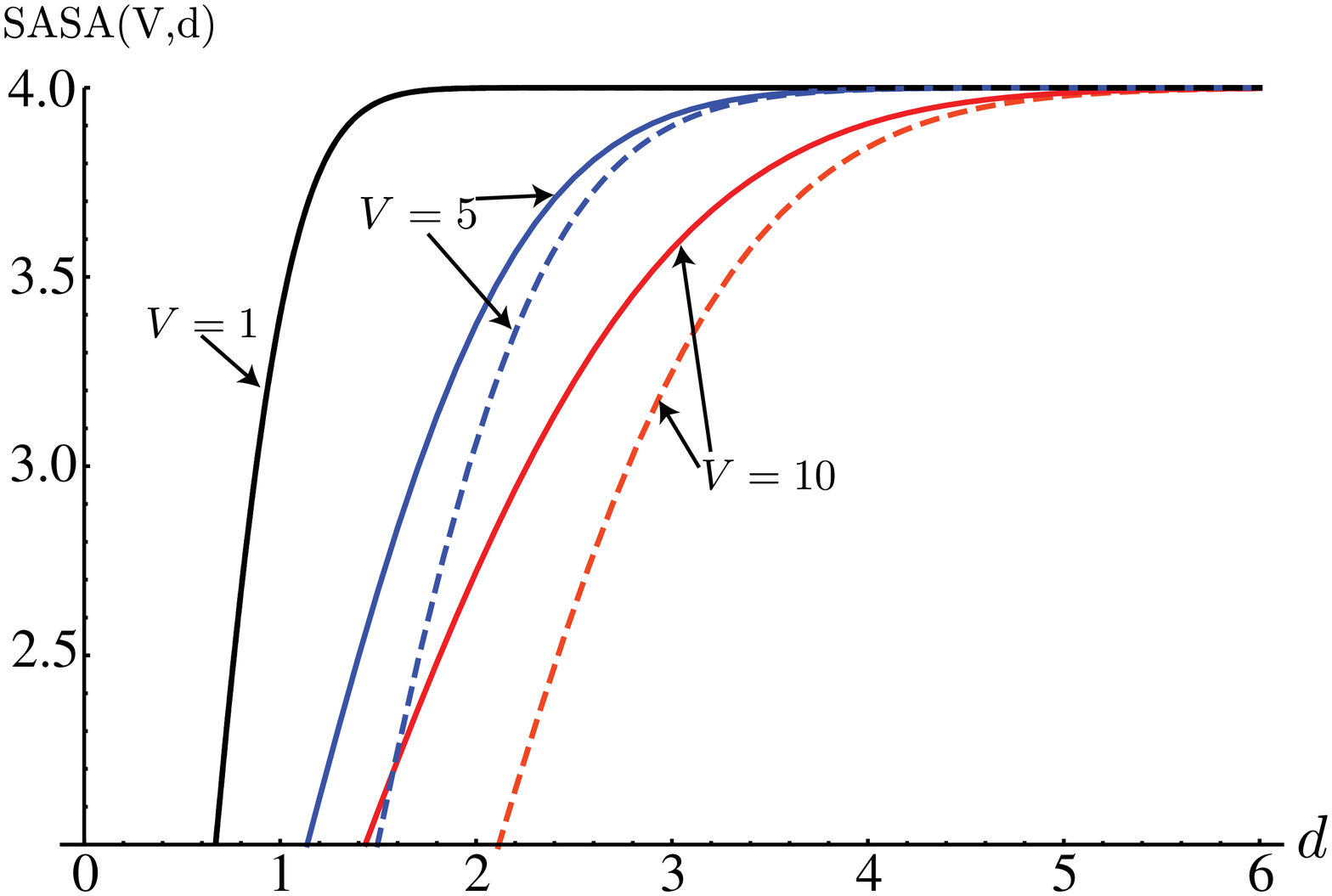,width=7cm,height=5.5cm}
\caption{Comparison between the SASA function associated with states $\rho^{(1)}_{cluster}$ (dashed line) and $\rho^{(2)}_{cluster}$ (solid line) for $V=1,5$ and 10.}
\label{comparison4}
\end{figure}

The protocol for state generation suggested in the previous Subsection is not the only possibility one has to generate linear cluster-like ETS. Indeed, along the lines of the scheme for tripartite GHZ-like ETS proposed in Fig.~\ref{generazione} {\bf (b)}, one can design situations where a suitably off-line prepared non-classical state of two modes is made available for interference at a set of beam splitters. This is the case for the procedure depicted in Fig.~\ref{generazione2} {\bf (b)}, where two thermal displaced states are fed into a cross-Kerr medium of large third-order non-linear rate $\chi$. A similar scheme has previously been put forward by Ba An and Hoa~\cite{others}. The inter-mode coupling model within the medium thus reads $\hat{H}_{\chi^{(3)}}=\hbar\chi\hat{a}^\dag_1\hat{a}_1\hat{a}^\dag_3\hat{a}_3$. By assuming an interaction time equal to $\pi/\chi$ and using the results of Ref.~\cite{mauro} one finds that the output modes are in state
\begin{equation}
\rho_\psi=\int d^2\alpha d^2\beta P^{\text{th}}_{\alpha}(V,d)P^{\text{th}}_{\beta}(V,d)\ket{\psi}_{13}\!\bra{\psi}
\end{equation}
with $\ket{\psi}_{13}=\ket{\alpha}_1|{s^+_\beta}\rangle_3+\ket{-\alpha}_1|{s^-_\beta}\rangle_3$. Interference of such a state with two vacuum-state modes  at $50:50$ beam splitters, together with local phase-shifting operations (not shown in Fig.~\ref{generazione2} {\bf (b)}), leave us with the state 
\begin{equation}
\rho^{(2)}_{cluster}{=}\int{d}^2\alpha\,{d}^2\beta{P}^{\text{th}}_\alpha(V,d){P}^{\text{th}}_\beta(V,d)\rho_c(\alpha',\beta')
\end{equation}
with $\rho_c(\alpha,\beta)=\ket{\text{C}(\alpha',\alpha',\beta',\beta')}_{1234}\!\bra{\text{C}(\alpha',\alpha',\beta',\beta')}$ and ${\alpha'=\alpha/\sqrt 2}$, ${\beta'=\beta/\sqrt 2}$. The crucial difference between $\rho^{(1)}_{cluster}$ and $\rho^{(2)}_{cluster}$ lies on the lack of independence between mode $1$ and $3$, as well as $2$ and $4$, which share the same amplitudes. This is the reason behind some difficulties in the full analytical treatment of the problem at hand, similar to those experienced in treating the situation described in Fig.~\ref{generazione} {\bf (b)}. We thus rely on a numerical analysis. Given that SASA inequality is optimal for linear cluster states, we concentrate on this case and compare the performance of the SASA function associated with $\rho^{(2)}_{cluster}$ and the behavior of Eq.~(\ref{sasafunc}) revealed in Fig.~\ref{C1}. Features qualitatively similar to those shown in the tripartite case are evident in Fig.~\ref{comparison4}, where the beam splitter-based approach is shown to be more advantageous as it allows for the violation of SASA inequality at slightly smaller values of $d$.

\section{Conclusions}
\label{conclu}

We have assessed the problem of revealing genuine multipartite non-locality in many-mode entangled states based on chaotic local components. We have identified in Svetlichny inequality and stabilizer-based arguments the crucial tools for our study, which has been performed by considering relevant instances of multipartite entangled states. In particular, we have considered tripartite GHZ and W-like ETS as well as quadripartite linear cluster states, revealing that the non-local properties of the class of states at hand can be faithfully revealed regardless of the thermal character of their local constituents and effectively counteracting the effects of detection inefficiency.  A small-scale experimental demonstration of our examples may be realized, not without some efforts related in particular to the technical difficulties in achieving strong optical nonlinearities. A three-mode GHZ ETS with $V\gtrsim1$ and $d\sim1.1$ can be generated, for instance, using the scheme in Fig.~\ref{generazione} {\bf (a)} using as an input state the superpositions of two coherent states with $d\sim1.6$, which were experimentally demonstrated in a recent seminal experiment~\cite{JacobNature}. This would pave the way towards the construction of a GHZ-like ETS and the test of tripartite nonlocality. The scaling-up of such technology towards the full implementation of the schemes here at hand could also take advantages from important progresses achieved in obtaining the demanding, yet necessary, strong nonlinearities.

Our work stands as a significant contribution to the current quest for quantum effects at the border of classicality, here embodied by large local temperatures and fuzzy measurement devices.

\acknowledgments
GMK is grateful to Steve Campbell for aid and support. MP thanks Gerardo Adesso and Alessio Serafini for discussions related to this work. FLS is partially supported by  CNPq (Brazil). His work is part of the Brazilian National Institute of Science and Technology of Quantum Information (INCT-IQ). 
HJ acknowledges support by Creative Research Initiatives (Center for
Macroscopic Quantum Control) of MEST/NRF, the World Class University (WCU)
program, the KOSEF grant funded by MEST/NRF (R11-2008-095-01000-0), and T.
J. Park Junior Faculty Fellowship. MP acknowledges support from  EPSRC (E/G004579/1) and hospitality from the Department of Physics and Astronomy, Seoul National University, where part of this work has been performed.

\end{document}